\documentclass[osajnl,twocolumn,showpacs,superscriptaddress,10pt]{revtex4-1}
\usepackage{amsmath,amssymb,graphicx}
\usepackage{ mathrsfs }
\usepackage{color}

\begin{document}

\title{{{Nonreciprocal unconventional photon blockade in a spinning optomechanical system}}}
\author{Baijun Li}
\affiliation{Key Laboratory of Low-Dimensional Quantum Structures
and Quantum Control of Ministry of Education, Department of
Physics and Synergetic Innovation Center for Quantum Effects and
Applications, Hunan Normal University, Changsha 410081, China}
\author{Ran Huang}
\affiliation{Key Laboratory of Low-Dimensional Quantum Structures
and Quantum Control of Ministry of Education, Department of
Physics and Synergetic Innovation Center for Quantum Effects and
Applications, Hunan Normal University, Changsha 410081, China}
\author{Xun-Wei Xu}
\affiliation{Department of Applied Physics, East China Jiaotong
University, Nanchang, 330013, China}
\author{Adam Miranowicz}\email{miran@amu.edu.pl}
\affiliation{Theoretical Quantum Physics Laboratory, RIKEN Cluster
for Pioneering Research, Wako-shi, Saitama 351-0198, Japan}
\affiliation{Faculty of Physics, Adam Mickiewicz University,
61-614 Pozna\'{n}, Poland}
\author{Hui Jing}\email{jinghui73@gmail.com}
\affiliation{Key Laboratory of Low-Dimensional Quantum Structures
and Quantum Control of Ministry of Education, Department of
Physics and Synergetic Innovation Center for Quantum Effects and
Applications, Hunan Normal University, Changsha 410081, China}

\begin{abstract}
We propose how to achieve quantum nonreciprocity via
unconventional photon blockade (UPB) in a compound device
consisting of an optical harmonic resonator and a spinning
optomechanical resonator. We show that, even with a very weak
single-photon nonlinearity, nonreciprocal UPB can emerge in this
system, i.e., strong photon antibunching can emerge only by
driving the device from one side, but not from the other side.
This nonreciprocity results from the Fizeau drag, leading to
different splitting of the resonance frequencies for the optical
counter-circulating modes. Such quantum nonreciprocal devices can
be particularly useful in achieving back-action-free quantum
sensing or chiral photonic communications.
\end{abstract}

\maketitle

\section{Introduction} \label{Int}

Photon blockade
(PB)~\cite{Tian92Quantum,Leonski94Possibility,Imamoglu97Strongly,Birnbaum05Photon,Muller15Coherent}, i.e., the generation of the first photon in a nonlinear cavity diminishes almost to zero the probability of generating another
photon in the cavity, plays a key role in
single-photon control for quantum technology applications nowadays~\cite{Gu17Microwave,Scarani09The,Buluta11Natural}. In experiments, PB
has been demonstrated in cavity-QED or circuit-QED systems~\cite{Birnbaum05Photon,Peyronel12Quantum,Muller15Coherent,Lang11Observation,Hoffman11Dispersive,Faraon08Coherent}. PB has also been predicted in
various nonlinear optical systems~\cite{Ferretti10Photon,Liao10Correlated,Miranowicz13Two}
and optomechanical (OM)
devices~\cite{Rabl11Photon,Nunnenkamp11Single,Liao13Photon,Xie16Photon,Zhai18mechanical}.
Conventional PB happens under the stringent condition of strong
single-photon nonlinearities, which is highly
challenging in practice.

To overcome this obstacle, coupled-resonator systems, with
destructive interferences of different dissipative
pathways~\cite{Leonski04Two,Miranowicz06Kerr,Liew10Single,Bamba11Origin}, have been proposed to achieve unconventional photon blockade
(UPB) even for arbitrarily weak
nonlinearities~\cite{Liew10Single,Bamba11Origin,Majumdar12Loss,Komar13Single,Savona13Unconventional,Xu13Antibunching,Ferretti13Optimal,Xu14Strong,Zhang14Optimal,Shen15Tunable,Flayac17Unconventional,Flayac17Nonclassical,Zhou18Photon,Snijders18Observation,Vaneph18Observation}.
UPB provides a
powerful tool to {generate optimally sub-Poissonian light} and also a
way to reveal quantum correlations in weakly nonlinear devices~\cite{Flayac17Unconventional,Flayac17Nonclassical}. Recently, UPB has been experimentally demonstrated in coupled
optical~\cite{Snijders18Observation} or superconducting
resonators~\cite{Vaneph18Observation}.

It should be stressed PB and UPB are very different
phenomena, thus also their nonreciprocal generalizations are also
different. Indeed PB refers to a process,
when a single photon is blocking the entry (or generation) of more
photons in a strongly nonlinear cavity. Thus, PB refers to state
truncation, also referred to as nonlinear quantum
scissors~\cite{Leonski01}. PB can be used as a source of single
photons, since the PB light is sub-Poissonian (or photon
antibunched) in second- and higher-orders, as characterized by the
correlation functions $g^{(n)}(0)<1$ for $n=2,3...$. By contrast
to PB, UPB refers to the light, which is optimally sub-Poissonian
in second order, $g^{2}(0)\approx 0$, and is generated in a
weakly-nonlinear system allowing for multi-path interference
(e.g., two linearly-coupled cavities, when one of them is also
weakly coupled to a two-level atom). Thus, PB and UPB are induced
by different effects: PB due to a large system nonlinearity and
UPB via multi-path interference assuming even an extremely-weak
system nonlinearity. Note that light generated via UPB can exhibit
higher-order super-Poissonian photon-number statistics,
$g^{(n)}(0)>1$ for some $n>2$. Thus, UPB is, in general,
\emph{not} a good source of single photons. This short comparison
of PB and UPB indicates that the term UPB, as coined in
Ref.~\cite{Carusotto13} and now commonly accepted, is fundamentally different
from PB, concerning their physical mechanisms and properties of their
generated light.

Here, we propose to achieve and control nonreciprocal UPB with
spinning devices.
{Nonreciprocal devices { allow for the flow of
light from one side but block it from the other. Thus, such
devices can be applied in noise-free quantum information signal
processing and quantum communication for cancelling interfering
signals~\cite{Manipatruni09Optical}.}} Nonreciprocal optical
devices have been realized in OM
devices~\cite{Manipatruni09Optical,Shen16Experimental,Bernier17Nonreciprocal},
Kerr
resonators~\cite{Cao17Experimental,Bino18Microresonator,Shi15Limitations},
thermo systems~\cite{Fan11An,Zhang18Thermal,Xia18Cavity}, devices
with temporal
modulation~\cite{Sounas17Non-reciprocal,Caloz18Electromagnetic},
and non-Hermitian
systems~\cite{Bender13Observation,Peng14Parity,Chang14Parity}. In
a very recent experiment~\cite{Maayani18Flying}, 99.6\% optical
isolation in a spinning resonator has been achieved based on the
optical Sagnac effect. However, these studies have mainly focused
on the classical regimes; that is, unidirectional control of
transmission rates instead of quantum noises. We also note that in
recent works, single-photon
diodes~\cite{Xia14Reversible,Tang18An,Scheucher16Quantum},
unidirectional quantum
amplifiers~\cite{Abdo14Josephson,Metelmann15Nonreciprocal,Malz18Quantum,Shen18Reconfigurable,Song18Direction},
and one-way quantum routers~\cite{Barzanjeh18Manipulating} have
been explored. In particular, nonreciprocal PB was predicted in a
Kerr resonator~\cite{Huang18Nonreciprocal} or a quadratic OM
system~\cite{Xu18arXiv}, which, however, relies on the
conventional condition of strong single-photon nonlinearity.
{These quantum nonreciprocal devices have potential
applications for quantum control of light in chiral and
topological quantum technologies~\cite{Lodahl17Chiral}.}
\begin{figure*}[hpbt]
\centerline{\includegraphics[width=0.98\textwidth]{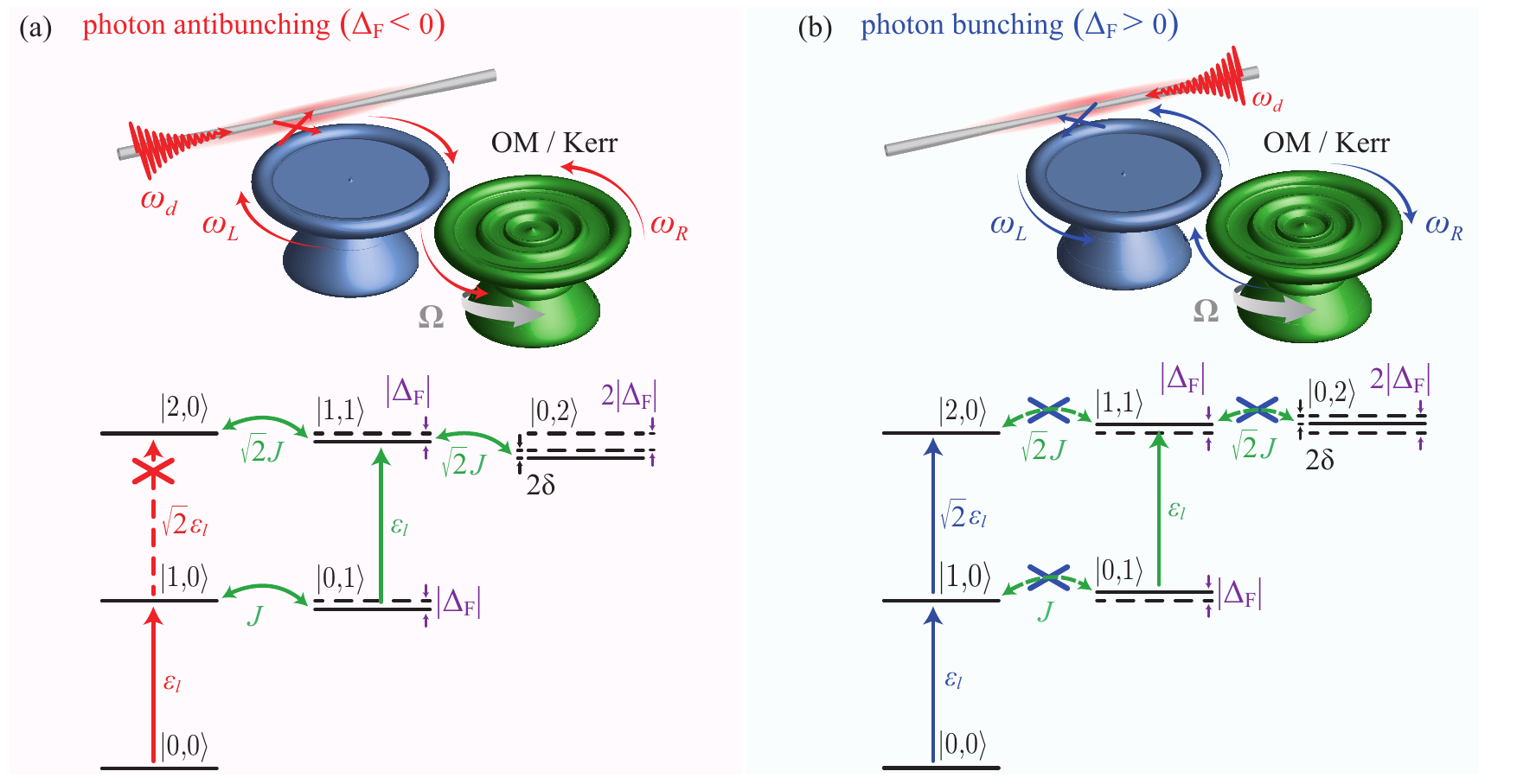}}
\caption{Nonreciprocal UPB in a coupled-resonator system. Spinning
the OM (Kerr-type) resonator results in different Fizeau drag
$\Delta_\mathrm{F}$ for the counter-circulating whispering-gallery
modes of the resonator. (a) By driving the system from the
left-hand side, the direct excitation from state $|1,0\rangle$ to
state $|2,0\rangle$ (red dotted arrow) will be forbidden by
destructive quantum interference with the other paths drawn by
green arrows, which leads to photon antibunching. (b) Photon
bunching occurs by driving the system from the right side, due to
the lack of the complete destructive quantum interference between
the indicated levels (drawn by crossed green dotted arrows). Here,
$\delta=g^2/\omega_m$ is the energy shift induced by the OM
nonlinearity.} \label{fig:FP}
\end{figure*}

We also note that coupled-cavity systems have been extensively studied in
experiments~\cite{Vaneph18Observation,Zhang18A,Konotop16Nonlinear,Ganainy18Non},
providing a unique way to achieve not only UPB, but also phonon
laser~\cite{Grudinin10Phonon,Jing14PT,Zhang18A}, slow light~\cite{Zhang18Loss}, and force
sensing~\cite{Liu16Metrology,Konotop16Nonlinear,Ganainy18Non}. Here we study nonreciprocal UPB in a coupled system
with an optical harmonic cavity and a spinning OM
resonator. We
find that, by the spinning of an OM resonator, UPB can emerge in a
nonreciprocal way even with a weak single-photon nonlinearity;
that is, strongly antibunched photons can emerge only by driving
the device from one side, but not the other side. Our work opens
up a new route to engineer quantum chiral UPB devices, which can
have practical applications in achieving, for example,
{photonic diodes or circulators, and nonreciprocal quantum
communications at the few-photon level.}

\section{Model and Solutions} \label{M and S}
We consider a compound system consisting of an optical harmonic
resonator (with the resonance frequency of the cavity field
$\omega_L$ and the decay rate $\kappa_L$) and a spinning
anharmonic resonator (with the resonance frequency of the cavity
field $\omega_R$ and the decay rate $\kappa_R$), as shown in
Fig.~\ref{fig:FP}. An external light is coupled into and out of
the resonator through a tapered fiber of frequency $\omega_d$ and
these two whispering-gallery-mode resonators are evanescently
coupled to each other with coupling strength
$J$~\cite{Spillane03Ideality}. Note that in the previous
proposal~\cite{Huang18Nonreciprocal}, requiring the strong Kerr
nonlinearity, $K\approx3\kappa$ (where $\kappa$ is the cavity
linewidth), is challenging for current experiments. Here we can
use experimentally feasible Kerr-nonlinear strength to realize
nonreciprocal PB; that is,
$K\approx0.04\kappa$~\cite{Vaneph18Observation}, which is two
orders of magnitude smaller than the former
work~\cite{Huang18Nonreciprocal}. Weak Kerr couplings can be
achieved in cavity-atom systems~\cite{Schmidt96Giant}, magnon
devices~\cite{Wang18Bistability}, and OM
systems~\cite{Gong09Effective} which we focus on here. We consider
a weakly OM coupling strength ($g\approx0.63\kappa$) in an
auxiliary cavity which is well within current experimental
abilities~\cite{Ding11Wavelength,Snijders16Purification,Enzian19Observation}.
In a spinning resonator, the refractive indices associated with
the clockwise ($\mathrm{+}$) and anticlockwise ($\mathrm{-}$)
optical modes are given as $n_{\pm} = n\left[1 \pm nv
(n^{-2}-1)/c\right]$, where $v=r\Omega$ is the tangential velocity
with the angular velocity $\Omega$ and radius
$r$~\cite{Maayani18Flying}. For light propagating in the spinning
resonator, optical mode experiences a Fizeau shift
$\Delta_\mathrm{F}$~\cite{Malykin00The}; that is, $\omega_R
\rightarrow \omega_R + \Delta_ \mathrm{F}$, with
\begin{align}\label{eq:Sagnacshift}
\Delta_\mathrm{F} &= \pm \frac{nr\Omega \omega_R}{c}
(1-\frac{1}{n^2}-\frac{\lambda}{n} \frac{\mathrm{d} n}{\mathrm{d}
\lambda}) = \pm \eta\Omega,
\end{align}
where $\omega_R=2\pi c/\lambda$ is the optical resonance frequency
for the nonspinning OM resonator, $c$ ($\lambda$) is the speed
(wavelength) of light in the vacuum, and $n$ is the refractive
index of the cavity. The dispersion term ${\mathrm{d}
n}/{\mathrm{d} \lambda}$, characterizing the relativistic origin
of the Sagnac effect, is relatively small in typical materials
($\sim1\%$)~\cite{Malykin00The,Maayani18Flying}.
For convenience, we always assume the counterclockwise
rotation of the resonator. Hence the $\pm \Delta_\mathrm{F}$
denote the light propagating against ($\Delta_\mathrm{F} > 0$) and
along ($\Delta_\mathrm{F} < 0$) the direction of the spinning OM
resonator, respectively.

In a rotating frame with respect to $H_0=\omega_d(a_L^{\dagger}a_L
+ a_R^{\dagger}a_R)$, the effective Hamiltonian of the system can
be written as (see Appendix~\ref{appendix A} for more details)
\begin{align}\label{eq:Ham1}
\mathcal{H}=&\ \hbar \Delta_L a_L^{\dagger} a_L + \hbar (\Delta_R + \Delta_ \mathrm{F}) a_R^{\dagger} a_R + \hbar \omega_m b^{\dagger} b \nonumber\\
&\ + \hbar J (a_L^{\dagger} a_R + a_R^{\dagger} a_L) + \hbar g
a_R^{\dagger} a_R (b^{\dagger} + b) \nonumber \\ &\ + i \hbar
\epsilon_d (a_L^{\dagger} -a_L),
\end{align}
where $a_L$ ($a_L^\dagger$) and $a_R$ ($a_R^\dagger$) are the
photon annihilation (creation) operators for the cavity modes of
the optical cavity (denoted by subscript $L$) and the OM cavity
(denoted by subscript $R$), respectively; $b$ ($b^{\dagger}$) is
the annihilation (creation) operator for the mechanical mode of
the OM cavity. The frequency detuning between the cavity field in
the left (right) cavity and the driving field is denoted by
$\Delta_K = \omega_K -\omega_d$ where  $K=L,R$; The parameter $J$
denotes the strength of the photon hopping interaction between the
two cavity modes; $g = \omega_R / r \left[\hbar/(2m \omega_m)
\right]^{1/2}$ describes the radiation-pressure coupling between
the optical and vibrative modes in the OM resonator with frequency
$\omega_m$ and effective mass $m$; $\epsilon_d=\sqrt{\kappa_L
P_\mathrm{in}/(\hbar\omega_d)}$ denotes the driving strength that
is coupled into the compound system through the optical fiber
waveguide with cavity loss rate $\kappa_L $ and driving power
$P_\mathrm{in}$.

The Heisenberg equations of motion of the system are then written
as:
\begin{align}\label{eq:L1}
\frac{\mathrm{d} }{\mathrm{d} t} q =&\ \omega_m p ,\nonumber \\
\frac{\mathrm{d} }{\mathrm{d} t} p =& - \omega_m q - g_b a_R^{\dagger}a_R - \frac{\gamma_m}{2}p + \xi ,\nonumber\\
\frac{\mathrm{d} }{\mathrm{d} t} a_L =& - \left(\frac{\kappa_L}{2} + i \Delta_L\right) a_L -i J a_R + \epsilon_d + \sqrt{\kappa_L} a_{L,\mathrm{in}} ,\nonumber\\
\frac{\mathrm{d} }{\mathrm{d} t} a_R =& - \left(\frac{\kappa_R}{2}
+ i \Delta_R'\right) a_R - i J a_L -ig_b q a_R  \nonumber \\&+
\sqrt{\kappa_L} a_{R,\mathrm{in}} ,
\end{align}
where $p$ and $q$ are dimensionless canonical position and
momentum with $p=i(b^{\dagger} - b) / \sqrt{2}$ and $q = (b +
b^{\dagger}) / \sqrt{2}$, respectively; $\Delta_R' = \Delta_R + \Delta_\mathrm{F}$, and $g_b =
\sqrt{2} g$; $\kappa_L=\omega_L/Q_L$
($\kappa_R=\omega_R/Q_R$) is the dissipation rate and $Q_L$
($Q_R$) is the quality factor of the left (right) cavity;
$\gamma_m=\omega_m/Q_M$ is damping rate with the quality factor
$Q_M$ of the mechanical mode. Moreover, $\xi$ is the zero-mean
Brownian stochastic operator, $\langle \xi(t) \rangle = 0$,
resulting from the coupling of the mechanical resonator with
corresponding thermal environment and satisfying the following
correlation function \cite{Ford88Quantum}:
\begin{align}\label{eq:Bc}
\langle \xi(t) \xi(t') \rangle = \frac{1}{2\pi} \int \mathrm{d}
\omega e^{-i \omega (t-t')} \Gamma_m(\omega),
\end{align}
where
\begin{align}\label{eq:Fc}
\Gamma_m(\omega) =  \frac{\omega
    \gamma_m}{2\omega_m} \left[1 + \coth\Big(\frac{\hbar\omega}{2k_BT} \Big) \right]
,
\end{align}
and $T$ is effective temperature of the environment of
the mechanical resonator and $k_B$ is the Boltzmann constant. The
annihilation operators $a_{L,{\mathrm{in}}}$ and
$a_{R,{\mathrm{in}}}$ are the input vacuum noise operators of the
optical cavity and the OM cavity with zero mean value,
respectively, i.e., $\langle a_{L,{\mathrm{in}}} \rangle = \langle
a_{R,{\mathrm{in}}} \rangle = 0$, and comply with
time-domain correlation
functions~\cite{Gardiner00Quantum,Walls94Quantum}:

\begin{align}\label{eq:gf}
\langle a_{K,{\mathrm{in}}}^{\dagger} (t)  a_{K,{\mathrm{in}}} (t')  \rangle =&\ 0,\nonumber \\
\langle a_{K,\mathrm{in}}(t)  a^{\dagger}_{K,\mathrm{in}} (t') \rangle  =&\ \delta (t-t'),
\end{align}
for $K = L, R$. Because the whole system interacts with a
low-temperature environment (here we consider 0.1~$\mathrm{mK}$),
we neglect the mean thermal photon numbers at optical frequencies
in the two cavities. In order to linearize the dynamics around the
steady state of the system, we expend the operators as the sum of
its steady-state mean values and a small fluctuations with zero
mean value around it; that is, $a_L = \alpha + \delta a_L$, $a_R =
\beta + \delta a_R$, $q = q_s + \delta q$, and $p = p_s + \delta
p$. {By neglecting higher-order terms, $\delta a_L^\dag\delta
a_L$, the {linearized} equations of the fluctuation
terms can be written as:
\begin{align}\label{eq:Ff1}
\frac{\mathrm{d} }{\mathrm{d} t} \delta q =&\  \omega_m \delta p , \nonumber\\
\frac{\mathrm{d} }{\mathrm{d} t} \delta p =& - \omega_m \delta q - g_b (\beta^* \delta a_R + \beta \delta a_R^{\dagger}) - \frac{\gamma_m}{2}\delta p + \xi , \nonumber\\
\frac{\mathrm{d} }{\mathrm{d} t} \delta a_L =& - \left(\frac{\kappa_L}{2} + i \Delta_L\right) \delta a_L -i J \delta a_R + \sqrt{\kappa_L}a_{L,\mathrm{in}}  , \nonumber\\
\frac{\mathrm{d} }{\mathrm{d} t} \delta a_R =& -
\left(\frac{\kappa_R}{2} + i\Delta_R' \right) \delta a_R - i J
\delta a_L - ig_b q_s\delta a_R \nonumber \\ &- i g_b \beta \delta
q+ \sqrt{\kappa_R} a_{R,\mathrm{in}}.
\end{align}
} {{These} equations can be solved in the
frequency domain (see Appendix~\ref{appendix B}). {In
particular, we find}}
\begin{align}\label{eq:Fs}
\delta a_L (\omega) =&\ E(\omega) a_{L,\mathrm{in}} (\omega) + F(\omega) a_{L,\mathrm{in}}^{\dagger} (\omega) + G(\omega) a_{R,\mathrm{in}} (\omega) \nonumber \\
&\ + H (\omega) a^{\dagger}_{R,\mathrm{in}} (\omega) + Q(\omega)
\xi (\omega),
\end{align}
where
\begin{align}\label{eq:Df}
E(\omega) =&\ \sqrt{\kappa_L}\, \frac{A_{1}(\omega)}{A_{5}(\omega)}  ,\nonumber\\
F(\omega) =& -\sqrt{\kappa_L}\, \frac{A_{2}(\omega)}{A_{5}(\omega)}  ,\nonumber\\
G(\omega) =&\ \sqrt{\kappa_R}\, \frac{A_{3}(\omega)}{A_{5}(\omega)}  , \nonumber\\
H(\omega) =& -\sqrt{\kappa_R}\, \frac{A_{4}(\omega)}{A_{5}(\omega)} ,\nonumber\\
Q(\omega) =& -i \frac{g_b \chi(\omega)}{\omega_m
A_{5}(\omega)}\left[\beta A_{3}(\omega) + \beta^* A_{4}
(\omega)\right] ,
\end{align}
and
\begin{align}\label{eq:Df1}
A_{1}(\omega) =& \left[\left(\frac{\kappa_R}{2} + i\omega\right) ^ 2 + \Delta_{R}''^2\right]V_1^{-}(\omega)\nonumber\\
&-g_b^4|\beta|^4 \left(\frac{\chi(\omega)}{\omega_m}\right)^2 V_1^{-}(\omega) + J^2 V_2^{+} ,\nonumber\\
A_{2}(\omega) =& -i J^2 g_b^2 \beta^2 \frac{\chi(\omega)}{\omega_m}  ,\nonumber\\
A_{3}(\omega) =& -i JV_1^{-}(\omega) V_2^{-} -i J^3  , \nonumber\\
A_{4}(\omega) =& -J g_b^2 \beta^2 \frac{\chi(\omega)} {\omega_m} V_1^{-}(\omega) ,\nonumber\\
A_{5}(\omega) =&\ V_1^{+}A_{1}(\omega) + i J A_{3} (\omega) ,
\end{align}
where we introduced the auxiliary functions:
\begin{align}\label{eq:Df2}
\Delta_R'' =&\ \Delta_R'  + g_b q_s - g_b^2 |\beta|^2 \chi(\omega),\nonumber\\
\chi(\omega) =&\ \omega_m^2 / (\omega_m^2 - \omega^2 + \frac{i \omega\gamma_m}{2}),\nonumber\\
V_1^\pm(\omega) =&\ \frac{\kappa_L}{2} \pm i(  \Delta_L  - \omega),\nonumber\\
V_2^\pm(\omega) =&\ {\frac{\kappa_R}{2} \pm i(  \Delta_R  -
\omega).}
\end{align}
\begin{figure*}[hpbt]
\centerline{\includegraphics[width=0.98\textwidth]{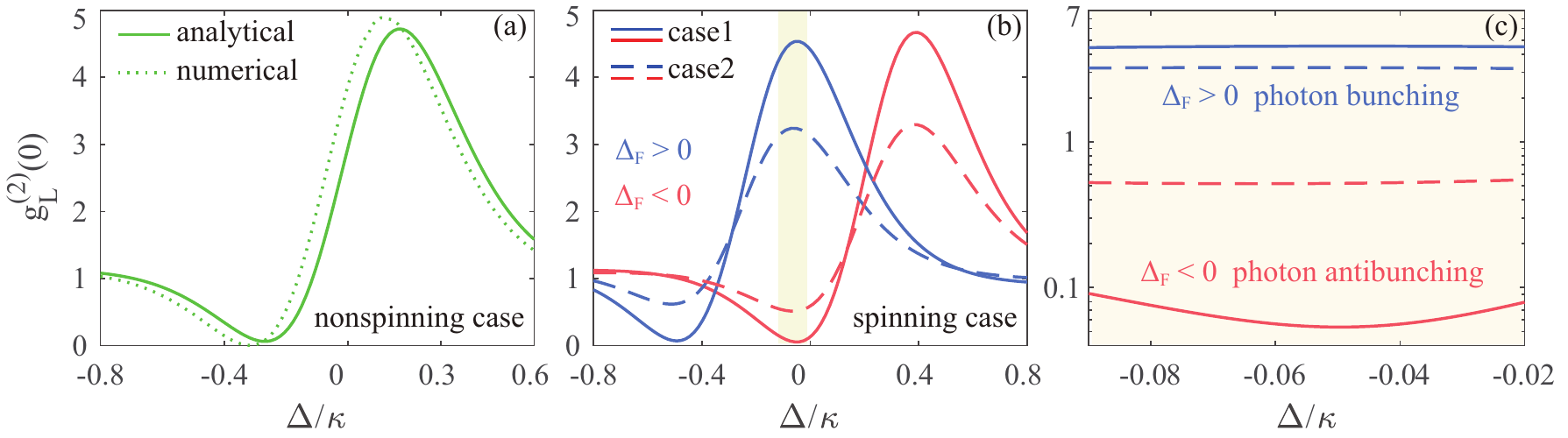}}
\caption{Correlation function $g_L^{(2)}(0)$ versus the
optical detuning $\Delta/\kappa$ (in units of the cavity loss rate
$\kappa_L = \kappa_R =\kappa$) with (a) $\Omega = 0$ and 
(b) $\Omega = 12~\mathrm{kHz}$, which is found numerically (solid
curves) and analytically (dotted curve). The PB can be generated
(red curves) or suppressed (blues curve) for different driving
directions, which can be seen more clearly in panel (c).
{The other parameters are: $g / \kappa= 0.63$,
$\omega_m /\kappa= 10$~\cite{Verhagen12Quantum}, $J / \kappa = 3$,
and $T = 0.1~\mathrm{mK}$ (case~1); $g / \kappa=
0.1$~\cite{Savona13Unconventional}, $\omega_m/ \kappa=
30$~\cite{Aspelmeyer14Cavity}, $J / \kappa = 20$, and $T =
1~\mathrm{mK}$ (case~2).}} \label{fig:FP1}
\end{figure*}

\section{Nonreciprocal Optical Correlations} \label{Q and C}
Now, we focus on the statistical properties of photons in optical
cavity, which are described quantitatively via normalized
zero-time delay second-order correlation function
$g_L^{(2)}(0)=\langle a_L^{\dagger2} a_L^{2}\rangle/\langle
a_L^{\dagger} a_L
\rangle^2$~\cite{Walls94Quantum,Xu13Antibunching}. By taking the
semiclassical approximation, i.e., $a_L = \alpha + \delta a_L$,
the correlation function $g_L^{(2)}(0)$ can be given
as~\cite{Xu13Antibunching}:
\begin{align}\label{eq:Gf}
g_L^{(2)}(0) =&\ \frac{|\alpha|^4 + 4|\alpha|^2 \mathcal{R}_{1}+
2\mathrm{Re} \left[\alpha^{*2}\mathcal{R}_{2} \right] +
\mathcal{R}_{3} }{\left(|\alpha|^2 + \mathcal{R}_{1} \right)^2},
\end{align}
where $\mathcal{R}_{1}=\langle \delta a_L^{\dagger}(t)\delta
a_L(t) \rangle$, $\mathcal{R}_{2}=\langle [\delta a_L(t)]^2\rangle$, and {{
$\mathcal{R}_{3}=\langle \delta a_L^{\dagger}(t)\delta
a_L^{\dagger}(t)\delta a_L(t)\delta a_L(t)
\rangle=2\mathcal{R}_{1}+|\mathcal{R}_{2}|^2$.}}

From Eq.~(\ref{eq:Fs}), the correlation between $\delta a_L(t)$
and $\delta a_L^{\dagger}(t)$ can be calculated as
\begin{align}\label{eq:Gfd}
   \langle \delta a_L^{\pm}(t) \delta a_L(t)\rangle =&\ \frac{1}{2\pi} \int_{-\infty}^{+\infty} X_{a_L^{\pm} a_L}d \omega,
\end{align}
where $\delta a_L^{+}(t) = \delta a_L^{\dagger}(t)$, $\delta
a_L^{-}(t) = \delta a_L(t)$, and
\begin{align}\label{eq:Gfd1}
X_{a_L^{\dagger} a_L} =&\ |Q(-\omega)|^2 \Gamma_m(-\omega) + |F(-\omega)|^2 + |H(-\omega)|^2,\nonumber\\
X_{a_La_L} =&\ Q(\omega) Q(-\omega) \Gamma_m(-\omega)
 + E(\omega) F(-\omega)\nonumber\\ &\ + G(\omega) H(-\omega).
\end{align}

To obtain more accurate results, we introduce the density operator
$\rho (t)$ and numerically calculate normalized zero-time delay
second-order correlation by the Lindblad master
equation~\cite{Johansson13Qutip}:
\begin{align}\label{eq:Mf}
\dot{\rho} =&\ \frac{1}{i\hbar} [\mathcal{H},\rho] +
\mathcal{L}[a_L](\rho)
 + \mathcal{L}[a_R](\rho) + \mathcal{L}[b](\rho) +\mathcal{L}[b^\dagger](\rho) ,
\end{align}
where $\mathcal{L}[o](\rho) = 2o\rho o^{\dagger} - o^{\dagger}
o\rho - \rho o^{\dagger}o$ are the Lindblad
superoperators~\cite{Walls94Quantum}, for $o = a_L$, $a_R$, $b$,
$b^{\dagger}$, and $\bar{n}_m = 1 / \left[\exp(\hbar \omega_m /
k_B T)-1 \right]$ is the mean thermal phonon numbers of the
mechanical mode at temperature $T$.

The second-order correlation function $g_L^{(2)}(0)$ is shown in
Fig.~\ref{fig:FP1} as function of the optical detuning
$\Delta/\kappa$ and the angular velocity $\Omega$. We assume
$\Delta_L = \Delta_R - \delta= \Delta$, $\kappa_L = \kappa_R =
\kappa$ and use experimentally feasible
parameters~\cite{Vahala03Optical,Peng14Parity,Teufel11Sideband,Ding11Wavelength,Verhagen12Quantum,Aspelmeyer14Cavity,Huet16Millisecond};
that is, $\lambda = 1550~\mathrm{nm}$, $Q_L = 3 \times 10^7$, $r =
0.3~\mathrm{mm}$, $n = 1.44$, $m =
5 \times 10^{-11}~\mathrm{kg}$,
$P_\mathrm{in} = 2 \times 10^{-17}~\mathrm{W}$. $Q_L$ is typically
$10^6-10^{12}$~\cite{Vahala03Optical,Aspelmeyer14Cavity,Huet16Millisecond},
$g$ is typically
$10^3-10^6~\mathrm{Hz}$~\cite{Ding11Wavelength,Verhagen12Quantum,Aspelmeyer14Cavity}
in optical microresonators, and
$g_L^{(2)}(0)\sim0.37$~\cite{Snijders18Observation,Vaneph18Observation}
was experimentally achieved. {$J$ can be adjusted by changing the distance of the double resonators~\cite{Zhang18A}.}
In a recent experiment,
autocorrelation measurements range from $g^{(2)}(0)=6\times 10^{-3}$
to $2$ have been achieved with average fidelity $0.998$ in a
photon-number-resolving detector~\cite{Hlousek18Accurate}.
Moreover, we set $\Omega=12~\mathrm{kHz}$, which is experimentally
feasible. The resonator with a radius of $r = 1.1~\mathrm{mm}$ can
spin at an angular velocity
$\Omega=6.6~\mathrm{kHz}$~\cite{Maayani18Flying}. By use of a
levitated OM system~\cite{Reimann18GHz,Ahn18Optically}, $\Omega$
can be increased even up to $\mathrm{GHz}$ values.
\begin{figure*}[hpbt]
\centerline{\includegraphics[width=0.98\textwidth]{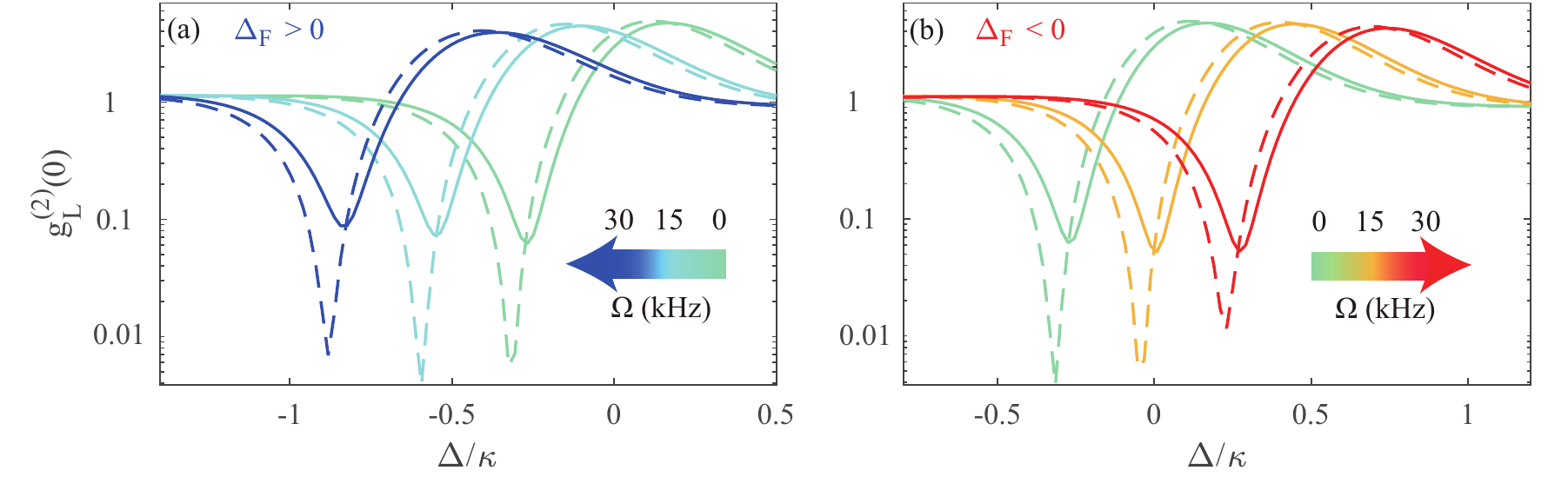}}
\caption{Correlation function $g_L^{(2)}(0)$ versus the
optical detuning $\Delta/\kappa$ (in units of cavity loss rate
$\kappa_L = \kappa_R =\kappa$) with various angular velocities
$\Omega$ by driving the device from the right side (a) or the
left-hand side (b). The dashed curves show our approximate
analytical results, given in Eq.~(\ref{eq:Gf}), while the solid
curves are our numerical solutions. The other parameters are the
same as those in Fig.~\ref{fig:FP1} (case~1).}
\label{fig:FP11}
\end{figure*}

Our analytical results agree well with the numerical
one. In the nonspinning-resonator case,
as shown in Fig.~\ref{fig:FP1}(a),
$g_L^{(2)}(0)$ is reciprocal regardless the direction of the
driving light, and always has a dip at
$\Delta/\kappa \approx -0.29$ and a peak
at $\Delta/\kappa \approx 0.166$, corresponding to strong photon
antibunching and photon bunching, respectively~\cite{Xu13Antibunching}. The physical
origin of strong photon antibunching is the destructive
interference between direct and indirect paths of two-photon excitations, i.e.,
\begin{align*}
&\left|1,0\right\rangle\stackrel{\sqrt{2}\epsilon_d}{\longrightarrow} \left|2,0\right\rangle,\\
&\left|1,0\right\rangle\stackrel{J}{\longrightarrow}\left|0,1\right\rangle
\stackrel{\epsilon_d}{\longrightarrow}\left|1,1\right\rangle
\stackrel{\sqrt{2}J}{\longrightarrow}\left|2,0\right\rangle.
\end{align*}

In contrast, for a spinning device, $g_L^{(2)}(0)$ exhibits giant
nonreciprocity, which can be seen in Fig.~\ref{fig:FP1}(b). The PB
can be generated, i.e., $g_L^{(2)}(0) \sim 0.06$, for
$\Delta_\mathrm{F}<0$, while significantly suppressed, i.e.,
$g_L^{(2)}(0) \sim 4.72$, for $\Delta_\mathrm{F} > 0$, which can
be seen more clearly in Fig.~\ref{fig:FP1}(c). The nonreciprocal
UPB induced by Fizeau light-dragging effect, with up to two orders
of magnitude difference of $g_L^{(2)}(0)$ for opposite directions,
can be achieved even with a weak nonlinearity and, to our
knowledge, has not been studied. {{Furthermore,
in Fig.~\ref{fig:FP1}(b), we use two sets of parameters for solid
(case~1) and dashed curves (case~2), respectively. It is seen
that nonreciprocity still exists in a parameter range closer to
the experiment.}}

{Since the anharmonicity of the system is very
small, destructive quantum interference (rather then the
anharmonicity) is responsible for observing strong photon
antibunching (referred to as UPB) and photon bunching (as referred
to photon-induced tunnelling) in the spinning devices as shown in
Fig.~\ref{fig:FP} and confirmed by our analytical calculations.
Note that the role of complete (incomplete) destructive quantum
interference is the same in both spinning and non-spinning UPB
systems, thus we refer to Ref.~\cite{Bamba11Origin}, where this
interference-based mechanism was first explained in detail.}
Spinning the OM resonator results in different Fizeau drag
$\Delta_\mathrm{F}$ for the counter-circulating whispering-gallery
modes of the resonator. By driving the system from the
left-hand side, the direct excitation from state $|1,0\rangle$ to
state $|2,0\rangle$ will be forbidden by the destructive quantum
interference with the indirect paths of two-photon excitations,
which leads to photon antibunching. In contrast, photon bunching
occurs by driving the system from the right side, due to the lack
of the complete destructive quantum interference between the
indicated levels~\cite{Reiter18Cooperative}. {Increasing the
angular velocity results in an opposing frequency shift of $\eta
\Omega$ for light coming from opposite directions. $g_L^{(2)}(0)$
also experiences linearly shifts with $\Omega$, but with different
directions for $\Delta_\mathrm{F}<0$ or $\Delta_\mathrm{F}>0$;
that is, we observe either a blueshift [see
Fig.~\ref{fig:FP11}(a)] or a redshift [see Fig.~\ref{fig:FP11}(b)]
with $\Delta_\mathrm{F}>0$ or $\Delta_\mathrm{F}<0$, respectively.
A highly-tunable nonreciprocal UPB device is thus achievable, by
flexible tuning of $\Omega$ and $\Delta/\kappa$. In addition,
since $g_L^{(2)}(0)$ is sensitive to $\Omega$, this may also
indicate a way for accurate measurements of velocity.}

\section{Optimal Parameters for Strong Antibunching} \label{O and P}

As discussed above, UPB can be generated nonreciprocally. In this
section, we analytically derive the optimal conditions of strong
antibunching.
{We apply here the method described in
Ref.~\cite{Bamba11Origin}, which is based on the evolution of a
complex non-Hermitian Hamiltonian, as given in Appendix~C. Thus,
our solution corresponds to only a semiclassical approximation of
the solution of the quantum master equation, given in
Eq.~(\ref{eq:Mf}), where the terms corresponding to quantum jumps
are ignored.}

Since the phonon states can be decoupled from the photon states by
using the unitary operator $U=\exp \left[-g(b^{\dagger} -
b)/\omega_m\right]$, the states of the system can be expressed as
$|\psi\rangle = |\varphi\rangle |\phi_m\rangle$, where
$|\varphi\rangle$ and $|\phi_\mathrm{m}\rangle$ are the photon
states and the phonon states, respectively. Under the weak-driving
condition, we make the ansatz~\cite{Bamba11Origin}
\begin{align}\label{eq:Sa}
|\varphi\rangle =&\ C_{00} |0,0\rangle + C_{10} |1,0\rangle +
C_{01} |0,1\rangle
 + C_{20} |2,0\rangle \nonumber\\&+ C_{11} |1,1\rangle + C_{02} |0,2\rangle  ,
\end{align}
and consider that $ C_{mn} \ll C_{m'n'} \ll C_{00}$ for $m+n=2$, $m'+n'=1$, and {the condition of $C_{20} = 0$,
the optimal conditions are given by fixing
$J$ and $\kappa$ (see Appendix~\ref{appendix C})}
\begin{align}\label{eq:Ov}
\Delta_\mathrm{opt} \approx&\ \frac{-a_3+\mathrm{sgn}(E) \sqrt{\lambda_1} - \sqrt{\lambda_2}}{4a_4},\nonumber\\
g_\mathrm{opt} =&\ \sqrt{-\frac{\omega_m\left[\Delta_\mathrm{opt}
(4\Delta_\mathrm{opt}^2 + 5\kappa^2 ) + \Delta_ \mathrm{F}
\lambda_3\right]}{2(2J^2-\kappa^2) + 2\Delta_
\mathrm{F}\lambda_{4}}},
\end{align}
{where $\mathrm{sgn}(E)$ is the signal function,
$a_3 =-96 \Delta_ \mathrm{F} \kappa$, {and $\lambda_{1,2}$,
which are defined in Appendix~\ref{appendix C}, are related to the
Fizeau drag $\Delta_ \mathrm{F}$}. Physically, this means
that the position of the minimum of $g^{(2)}_L(0)$ is determined by
the detuning between the two cavity fields. Thus, $\Delta_
\mathrm{F}$ can lead to a shift of the minimum of $g^{(2)}_L(0)$ to
achieve nonreciprocity.}

\begin{figure}[hpbt]
    \centerline {\includegraphics[width=0.48\textwidth]{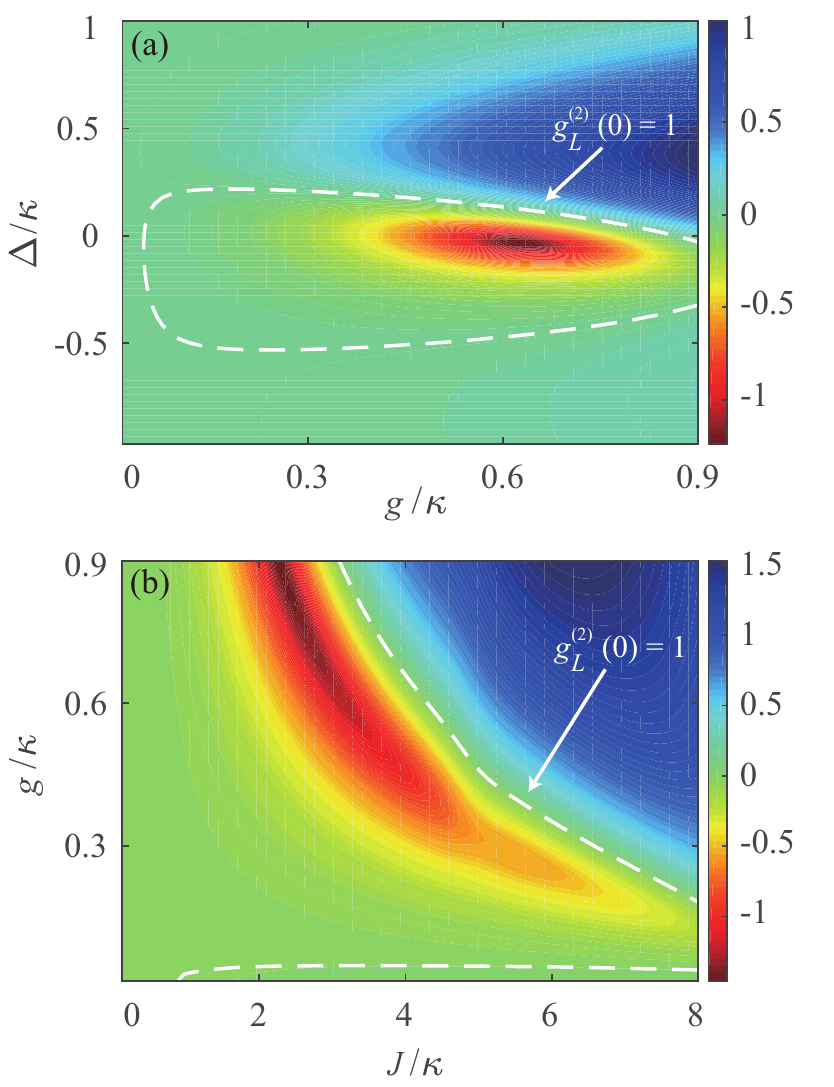}}
    \caption{Correlation function $g_L^{(2)}(0)$ in logarithmic
        scale [i.e., $\log_{10}g_L^{(2)}(0)$] versus: (a) the
        radiation-pressure coupling $g/\kappa$ (in units of cavity loss
        rate $\kappa = \kappa_L = \kappa_R$) and the optical detuning
        $\Delta/\kappa$; and (b) the coupling strength of the resonators
        $J/\kappa$ and the radiation-pressure coupling $g/\kappa$ for the
        optical detuning $\Delta/\kappa=-0.05$. The angular
        velocity is $\Omega = 12~\mathrm{kHz}$ and the white dotted curve
        corresponds to $g_L^{(2)}(0)=1$. The
        other parameters are the same as those in Fig.~\ref{fig:FP11}.}
    \label{fig:FP12}
\end{figure}
\begin{figure}[hpbt]
    \centerline {\includegraphics[width=0.48\textwidth]{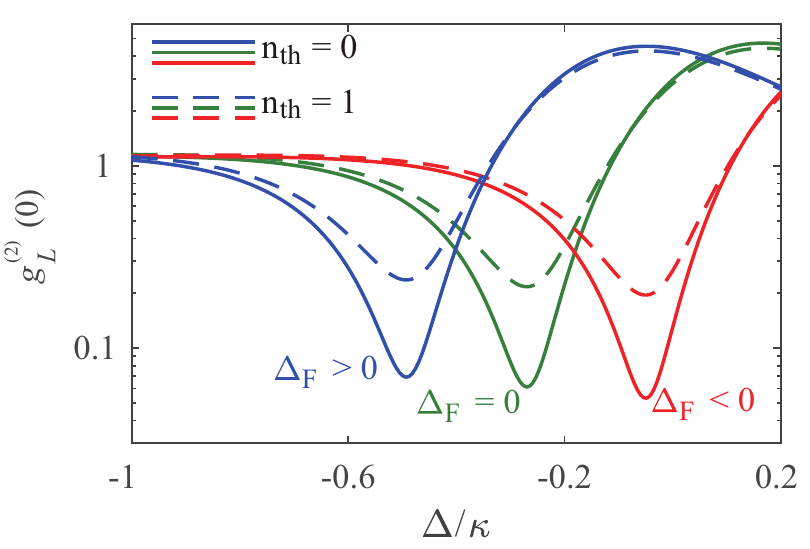}}
    \caption{Correlation function $g_L^{(2)}(0)$ versus
        the optical detuning $\Delta/\kappa$ (in units of cavity loss rate
        $\kappa_L = \kappa_R =\kappa$) with varied mean thermal phonon
        numbers $\mathrm{n_{\mathrm{th}}}$ for various angular velocities
        $\Omega$, and the resulting Fizeau shifts $\Delta_\mathrm{F}$. The
        other parameters are the same as those in Fig.~\ref{fig:FP12}.}
    \label{fig:FP13}
\end{figure}

\begin{figure*}[hpbt]
\centerline {\includegraphics[width=0.98\textwidth]{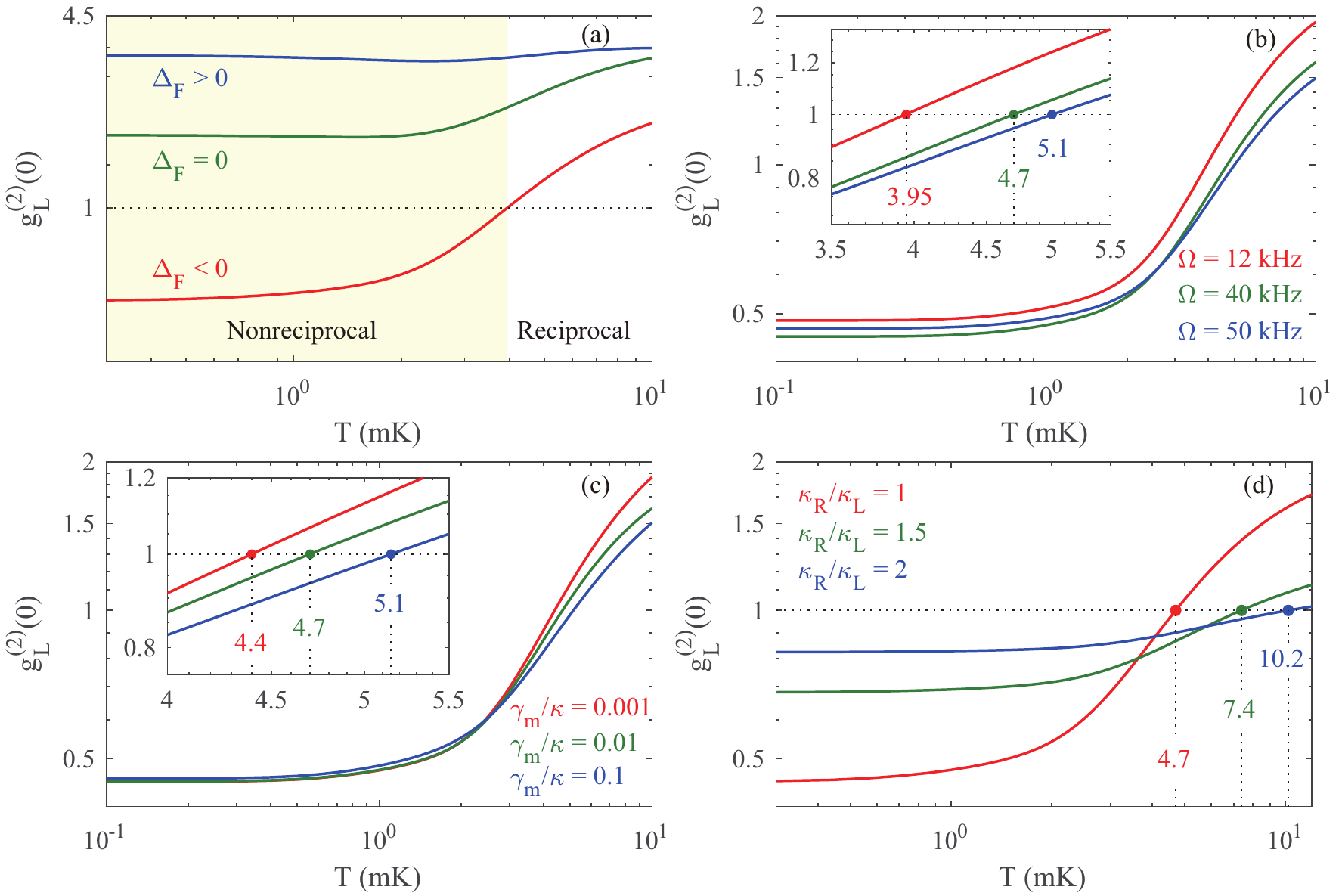}}
\caption{{(a) Correlation function
$g_L^{(2)}(0)$ {versus the effective temperature $T$ of the
environment of the mechanical resonator} for three values of the
Fizeau shifts $\Delta_\mathrm{F}$ ($\Delta_\mathrm{F}>0$,
$\Delta_\mathrm{F}=0$, and $\Delta_\mathrm{F}<0$) for the optimal
values of $\Delta_\mathrm{opt}$ and $g_\mathrm{opt}$. {The
other parameters are set the same as in case~2 in
Fig.~\ref{fig:FP1}. Moreover,} the correlation function
$g_L^{(2)}(0)$ versus $T$ for various values of: (b) the
spinning frequency, (c) the mechanical decay, and (d) the cavity
decay, {assuming that the device is driven from the left-hand
side and the optical detuning is fixed at the optimal values.}}}
\label{fig:FP16}
\end{figure*}

In order to visualize UPB more clearly, we show the contour plots
of $g_L^{(2)}(0)$ in logarithmic scale [i.e.,
$\log_{10}g_L^{(2)}(0)$] as function of $g/\kappa$ and
$\Delta/\kappa$ in Fig.~\ref{fig:FP12}(a). By fixing
$\Delta/\kappa=-0.05$, we obtain the function of $g_L^{(2)}(0)$ in
logarithmic scale versus the coupling strength of the resonators
$J/\kappa$ and $g/\kappa$, as shown in Fig.~\ref{fig:FP12}(b).
These plots show that strong photon antibunching occurs exactly at
the values predicted from our analytical calculations in
Eq.~(\ref{eq:Ov}). Moreover, by computing $g_L^{(2)}(0)$ as the
function of $\Delta/\kappa$ and $\Omega$ with different mean
thermal phonon numbers $\mathrm{n_{\mathrm{th}}}$, as shown in
Fig.~\ref{fig:FP13}, we confirm that rotation-induced
nonreciprocity can still exist by considering thermal phonon
noises. {We note that thermal phonons greatly
affect the correlation $g_L^{(2)}(0)$ of photons and tend to
destroy photon blockade. Thus, to show this effect, in
Fig.~\ref{fig:FP16}(a), we plot the correlation $g_L^{(2)}(0)$ as
a function of temperature $T$ for various Fizeau shifts.
{We see that nonreciprocal UPB can be observed below the
critical temperature $T_0\approx4~\mathrm{mK}$ ($5~\mathrm{mK}$)
for the spinning frequency $\Omega=12~\mathrm{kHz}$
($\Omega=50~\mathrm{kHz}$) [see Fig.~\ref{fig:FP16}(b)]. By
further increasing the optical dissipation of the optomechanical
cavity, as shown in Fig.~\ref{fig:FP16}(d), the critical
temperature $T_0$ can reach the value of $10~\mathrm{mK}$.}}

Finally, we note that a state (generated via UPB or another
effect) with vanishing (or almost vanishing) second-order
photon-number correlations, $g^{(2)}(0)\approx 0$, is \emph{not}
necessarily a good single-photon source, i.e., the state might not
be a (partially-incoherent) superposition of \emph{only} the
vacuum and single-photon states. A good single-photon source is
characterized not only by $g^{(2)}(0)\approx 0$, but also by
vanishing higher-order photon-number correlation functions,
$g^{(n)}(0)\approx 0$ for $n>2$. In UPB, $g^{(n)}(0)$ for $n>2$
can be greater than $g^{(2)}(0)\approx 0$, or even greater than
1~\cite{Radulaski17}. Indeed a standard
analytical method for analyzing UPB, as proposed by Bamba \emph{et
al.}~\cite{Bamba11Origin} and applied here, is based
on expanding the wave function $|\varphi\rangle$ of a
two-resonator system in power series $|\varphi\rangle=\sum
C_{n,m}|n,m\rangle$ up to the terms $C{n,2-n}$ ($n=0,1,2$) only,
as given in Eq.~(\ref{eq:Sa}). To obtain the optimal system
parameters, which minimize $g^{(2)}(0)$ in UPB, this method
requires to set $C_{2,0}=0$ as set in Appendix C. Actually, the
same expansion of $|\varphi\rangle$ and same ansatz are made in
Ref.~\cite{Bamba11Origin}. These assumptions imply that
higher-order correlation functions $g^{(n)}(0)$ with $n=3,4...$
vanish too. However, the truncation of the above expansion at the
terms $C_{n,2-n}$ is often not justified for a system exhibiting
UPB. Indeed, we find parameters for our system, for which
$g^{(2)}(0)\approx 0$ and simultaneously $g^{(3)}(0)>1$. We have confirmed this
by a precise numerical calculation of the steady states
of our system based on the non-Hermitian Hamiltonian, given in
Eq.~(\ref{eq:Ke1}), in the Hilbert space larger than $4\times4$.

\section{Conclusions} \label{C}

In summary, we studied nonreciprocal UPB in a system consisting of
a purely optical resonator and a spinning OM resonator. Due to the
interference between two-photon excitations paths and the Sagnac
effect, UPB can be generated nonreciprocally in our system; that
is, UPB can occur when the system is driven from one direction but
not from the other, even under the weak OM interactions. The
optimal conditions for one-way UPB were given analytically.
Moreover, we found this quantum nonreciprocity can still exist by
considering thermal phonon noises.

{Concerning a possible experimental implementation of
nonreciprocal UP, it is worth noting that UPB for non-spinning
devices has already been demonstrated experimentally in two recent
works~\cite{Snijders18Observation,Vaneph18Observation}. A number of experiments (including the very recent
work~\cite{Maayani18Flying}) have shown non-reciprocal quantum effects in spinning
devices. So the main experimental task for achieving
non-reciprocal UPB in a spinning device would be to combine
experimental setups of, e.g., Refs.~\cite{Snijders18Observation,Vaneph18Observation,Maayani18Flying} into a single
spinning UPB setup.}

Our proposal provides a feasible method to control the behavior of
one-way  photons, with the potential applications in achieving,
e.g., photonic diodes or circulators, quantum chiral
communications, and nonreciprocal light engineering in deep
quantum regime.

\appendix
\section{Derivation of Effective Hamiltonian} \label{appendix A}
The coupled system can be described by the following Hamiltonian
\begin{align}\label{eq:hamun}
H =&\ H_0 + H_\mathrm{in} + H_\mathrm{dr}, \nonumber \\
H_0 =&\ \hbar \omega_L a_L^{\dagger} a_L + \hbar (\omega_R + \Delta_ \mathrm{F}) a_R^{\dagger} a_R + \hbar \omega_m b^{\dagger} b, \nonumber \\
H_ \mathrm{in} =&\ \hbar J (a_L^{\dagger} a_R + a_R^{\dagger} a_L) + \hbar g a_R^{\dagger} a_R (b^{\dagger} + b), \nonumber \\
H_\mathrm{dr} =&\ i \hbar \epsilon_d (a_L^{\dagger} e^{-i \omega_d
t} - a_L e^{i \omega_d t}),
\end{align}
where $a_L$ ($a_L^\dagger$) and $a_R$ ($a_R^\dagger$) are the
photon annihilation (creation) operators for the cavity modes of
the optical cavity (denoted by subscript $L$) and the OM cavity
(denoted by subscript $R$), respectively; $b$ ($b^{\dagger}$) is
the annihilation (creation) operator for the mechanical mode of
the OM cavity. The frequencies of the cavity fields are denoted by
$\omega_L$ and $\omega_R$. $J$ is the coupling strength between
the two resonators, $g = \omega_R/r \left[\hbar/(2m \omega_m)
\right]^{1/2}$ is the OM coupling strength between the optical
mode and the mechanical mode in the OM cavity,
$\epsilon_d=\sqrt{\kappa_L P_\mathrm{in}/(\hbar\omega_d)}$
denotes the driving strength which is coupled into the compound
system through the optical fiber waveguide.

Using the unitary operator $U=\exp \left[-g(b^{\dagger} -
b)/\omega_m\right]$ to Hamiltonian~(\ref{eq:hamun}), we obtain a
Kerr-type one~\cite{Gong09Effective}
\begin{align}\label{eq:Ham2}
H_ \mathrm{eff} =&\ U^\dagger H U \nonumber\\
=&\ \hbar \omega_L a_L^{\dagger} a_L + \hbar (\omega_R + \Delta_\mathrm{F}) a_R^{\dagger} a_R - \hbar \delta (a_R^{\dagger} a_R)^2 \nonumber \nonumber \\ &\ + \hbar J \left[a_L^{\dagger} a_R e^{-\delta (b^{\dagger} - b)} + a_L a_R^{\dagger} e^{\delta (b^{\dagger} - b)}\right] \nonumber \\
&\ + i \hbar \epsilon_d (a_L^{\dagger} e^{-i \omega_d t} - a_L
e^{i \omega_d t}),
\end{align}
where $\delta = g^2 / \omega_m$. Under the conditions,
$g/\omega_m\ll1$ and $J<\omega_m/2$, the
Hamiltonian~(\ref{eq:Ham2}) can be read as
\begin{align}\label{eq:Ham3}
H_ \mathrm{eff}' =&\ \hbar \omega_L a_L^{\dagger} a_L + \hbar (\omega_R + \Delta_\mathrm{F}) a_R^{\dagger} a_R - \hbar \delta (a_R^{\dagger} a_R)^2 \nonumber \nonumber \\ &\ + \hbar J \left[a_L^{\dagger} a_R + a_L a_R^{\dagger}\right] \nonumber \\
&\ + i \hbar \epsilon_d (a_L^{\dagger} e^{-i \omega_d t} - a_L
e^{i \omega_d t}).
\end{align}

\section{The Fourier Analysis of Fluctuation Terms } \label{appendix B}
According to the Heisenberg equations of motion of
Hamiltonian~(\ref{eq:Ham1}), and using semiclassical approximation
method, i.e., $a_L = \alpha + \delta a_L$, $a_R = \beta + \delta
a_R$, $q = q_s + \delta q$, and $p = p_s + \delta p$, the
steady-state values of the system satisfy the following equations:
\begin{align}\label{eq:Sf}
0=&\left(\frac{\kappa_L}{2} + i \Delta_L\right) \alpha + i J \beta -\epsilon_d,\nonumber\\
0=&\left[\frac{\kappa_R}{2} + i(\Delta_R' + g_b q_s) \right]\beta - i J \alpha,\nonumber\\
0=&\ \omega_m q_s -g_b |\beta|^2.
\end{align}
Then we obtain
\begin{align}\label{eq:Sf1}
b_3 q_s^3 + b_2 q_s^2  +b_1 q_s + b_0=0,
\end{align}
where

\begin{align}\label{eq:Sf2}
b_0 =&\ g_b J^2 \epsilon_d^2 ,\nonumber\\
b_1 =&\  \omega_m \left(\frac{\kappa_L\kappa_R}{4} + J^2\right)^2 + \omega_m \left(\frac{\kappa_L\Delta_R'}{2}+ \frac{\kappa_R\Delta_L}{2}\right)^2  \nonumber \\ &\ -\omega_m\Delta_L\Delta_R'\left(\frac{\kappa_L\kappa_R}{2} + 2J^2 -\Delta_L\Delta_R'\right),\nonumber\\
b_2 =&\ 2\omega_m g_b\left[\frac{\kappa_L^2\Delta_R'}{4} +  \Delta_L (\Delta_L\Delta_R' - J^2) \right] ,\nonumber \\
b_3 =&\ \omega_m g_b^2\left(\frac{\kappa_L^2 }{4} +
\Delta_L^2\right) .
\end{align}

The fluctuation terms of the system can be written as:
\begin{align}\label{eq:Ff}
\frac{\mathrm{d} }{\mathrm{d} t} \delta q =&\  \omega_m \delta p , \nonumber\\
\frac{\mathrm{d} }{\mathrm{d} t} \delta p =& - \omega_m \delta q - g_b (\beta^* \delta a_R + \beta \delta a_R^{\dagger}) - \frac{\gamma_m}{2}\delta p + \xi , \nonumber\\
\frac{\mathrm{d} }{\mathrm{d} t} \delta a_L =& - \left(\frac{\kappa_L}{2} + i \Delta_L\right) \delta a_L -i J \delta a_R + \sqrt{\kappa_L}a_{L,\mathrm{in}}  , \nonumber\\
\frac{\mathrm{d} }{\mathrm{d} t} \delta a_R =& -
\left(\frac{\kappa_R}{2} + i\Delta_R' \right) \delta a_R - i J
\delta a_L - ig_b q_s\delta a_R \nonumber \\ &- i g_b \beta \delta
q+ \sqrt{\kappa_R} a_{R,\mathrm{in}} ,
\end{align}
where we have neglected higher-order terms, $\delta a_L^\dag\delta
a_L$. Here, the steady-state mean value $q_s$ is numerically
solved from Eqs.~(\ref{eq:Sf1}) and (\ref{eq:Sf2}).

By introducing the Fourier transform to the fluctuation equations,
we find:
\begin{align}\label{1}
i\omega \delta a_L(\omega)=&- \left(\frac{\kappa_L}{2}+i\Delta_L\right)\delta a_L(\omega)-iJ\delta a_R(\omega) \nonumber\\&+\sqrt{\kappa_L}a_{L,\mathrm{in}}(\omega),\nonumber\\
i\omega \delta a_R(\omega)=&-\left(\frac{\kappa_R}{2}+i\Delta_R'''\right)\delta a_L(\omega) -iJ\delta a_R(\omega) \nonumber\\&-ig_b\beta \delta q (\omega)+\sqrt{\kappa_R}a_{R,\mathrm{in}}(\omega),\nonumber\\
i\omega\delta q(\omega) =&\ \omega_m \delta p (\omega),\nonumber\\
i\omega \delta p (\omega)=&-\omega_m \delta q (\omega)
-g_b\left[\beta^*\delta a_R (\omega) + \beta \delta a_R^\dagger
(\omega)\right]\nonumber\\ &- \frac{\gamma_m}{2}\delta p(\omega) +
\xi(\omega),
\end{align}
where $\Delta_R'''=\Delta_R'+g_bq_s$, then we obtain
\begin{align}\label{2}
\delta q(\omega) =& -g_b\beta^*\chi(\omega)\delta
a_R(\omega)-g_b\beta\chi(\omega)\delta a_R^\dagger(\omega)
\nonumber\\&+ \chi(\omega)\xi(\omega),
\end{align}
where
\begin{align}
\chi(\omega)=\frac{\omega_m}{\omega_m^2-\omega^2+i\omega\gamma_m/2}.
\end{align}
Substituting Eq.~(\ref{2}) into Eq.~(\ref{1}), we have
\begin{align}\label{3}
M(\omega)\delta a_R(\omega)=&\ ig_b^2\beta^2\chi(\omega)\delta a_R^\dagger(\omega)-ig_b\beta\chi(\omega)\xi(\omega)\nonumber\\
&\ -iJ\delta_L(\omega)+\sqrt{\kappa_R}a_{R,\mathrm{in}}(\omega),
\end{align}
where
\begin{align}
M(\omega)=\frac{\kappa_R}{2}+i\omega+i\Delta_R'''-i|\beta|^2g_b^2\chi(\omega).
\end{align}

According to Eq.~(\ref{1}), we obtain
\begin{align}\label{4}
i\omega \delta a_L^\dagger(\omega)=&-\left(\frac{\kappa_L}{2}-i\Delta_L\right)\delta a_ L^\dagger(\omega)+iJ\delta a_R^\dagger(\omega)\nonumber\\&+\sqrt{\kappa_L}a_{L,\mathrm{in}}^\dagger(\omega),\nonumber\\
i\omega \delta a_R^\dagger(\omega)=&-\left(\frac{\kappa_R}{2}-i\Delta_R'''\right)\delta a_R^\dagger(\omega) +iJ\delta a_R^\dagger(\omega) \nonumber\\&+ig_b\beta \delta q^\dagger (\omega)+\sqrt{\kappa_R}a_{R,\mathrm{in}}^\dagger(\omega),\nonumber\\
i\omega\delta q^\dagger(\omega) =&\ \omega_m \delta p^\dagger (\omega),\nonumber\\
i\omega \delta p^\dagger (\omega)=&-\omega_m \delta q^\dagger
(\omega)-g_b\left[\beta \delta a_R^\dagger (\omega)+\beta^*\delta
a_R (\omega)\right]\nonumber\\ &- \frac{\gamma_m}{2}\delta
p^\dagger + \xi^\dagger(\omega),
\end{align}
then we have
\begin{align}\label{5}
N(\omega)\delta a_R(\omega)=&-ig_b^2\beta^{*2}\chi(\omega)\delta a_R^\dagger(\omega)+ig_b\beta^*\chi(\omega)\xi^\dagger(\omega)\nonumber\\
&+iJ\delta
a_L^\dagger(\omega)+\sqrt{\kappa_R}a_{R,\mathrm{in}}^\dagger(\omega),
\end{align}
where
\begin{align}
N(\omega)=\frac{\kappa_R}{2}+i\omega-i\Delta_R'''+i|\beta|^2g_b^2\chi(\omega).
\end{align}

From Eq.~(\ref{4}), we have
\begin{align}\label{6}
V(\omega)\delta a_L^\dagger(\omega)=iJ\delta a_R^\dagger(\omega)
+\sqrt{\kappa_L}  a_{L,\mathrm{in}}^\dagger(\omega),
\end{align}
where $V(\omega)=\kappa_L/2+i\omega -i\Delta_L$. Substituting
Eq.~(\ref{6}) into Eq.~(\ref{5}), we find
\begin{align}\label{7}
T(\omega)\delta a_R^\dagger(\omega)=& -i\chi(\omega)g_b^2\beta^{*2}V(\omega)\delta a_R(\omega) \nonumber\\
&+i\chi(\omega)g_b\beta^{*}V(\omega)\xi^\dagger(\omega) \nonumber\\
&+iJ\sqrt{\kappa_L}a_{L,\mathrm{in}}^\dagger(\omega)
+\sqrt{\kappa_R}V(\omega)a_{R,\mathrm{in}}^\dagger(\omega),
\end{align}
where $T(\omega)= N(\omega)V(\omega)+J^2$. Substituting
Eq.~(\ref{7}) into Eq.~(\ref{3}), we obtain
\begin{align}\label{8}
F_R(\omega)\delta a_R(\omega)=&-\chi^2(\omega)g_b^3\beta|\beta|^2V(\omega)\xi^\dagger(\omega)\nonumber\\
&-ig_b\beta\chi(\omega)T(\omega)\xi(\omega)\nonumber\\
&-Jg_b^2\beta^2\chi(\omega)\sqrt{\kappa_L}a_{L,\mathrm{in}}^\dagger\nonumber\\
&+ig_b^2\beta^2\chi(\omega)\sqrt{\kappa_R}V(\omega)a_{R,\mathrm{in}}^\dagger(\omega)\nonumber\\
&-iJT(\omega)a_{L,\mathrm{in}}
-\sqrt{\kappa_R}T(\omega)a_{R,\mathrm{in}},
\end{align}
where the auxiliary function are
$F_R(\omega)=M(\omega)T(\omega)-\chi^2(\omega)V(\omega)
g_b^4|\beta|^4$. Substituting Eq.~(\ref{8}) into Eq.~(\ref{1}), we
have
\begin{align}\label{9}
F_L(\omega)\delta a_L(\omega)=&\ iJ\chi^2(\omega)g_b^3\beta|\beta|^2V(\omega)\xi^\dagger(\omega)\nonumber\\
&\ -g_b\beta\chi(\omega)JT(\omega)\xi(\omega)\nonumber\\
&\ +iJ^2g_b^2\beta^2\chi(\omega)\sqrt{\kappa_L}a_{L,\mathrm{in}}^\dagger\nonumber\\
&\ +Jg_b^2\beta^2\chi(\omega)\sqrt{\kappa_R}V(\omega)a_{R,\mathrm{in}}^\dagger(\omega)\nonumber\\
&\ -iJ\sqrt{\kappa_R}T(\omega)a_{R,\mathrm{in}}\nonumber\\
&\
-\sqrt{\kappa_L}\left[M(\omega)T(\omega)-U(\omega)\right]a_{L,\mathrm{in}},
\end{align}
where
\begin{align}
F_L(\omega)=&\left[M(\omega)T(\omega)-U(\omega)\right]V_1(\omega)+J^2T(\omega),\nonumber\\
U(\omega)=&-\chi(\omega)^2 g_b^4|\beta|^4(i\omega+\frac{\kappa_L}{2}-i\Delta_L),\nonumber\\
V_1(\omega)=&\ \frac{\kappa_L}{2}+ i\omega+i\Delta_L.
\end{align}
Then we find
\begin{align}\label{10}
\delta a_L (\omega) =&\ E(\omega) a_{L,\mathrm{in}} (\omega) + F(\omega) a_{L,\mathrm{in}}^{\dagger} (\omega) + G(\omega) a_{R,\mathrm{in}} (\omega) \nonumber \\
&\ + H (\omega) a^{\dagger}_{R,\mathrm{in}} (\omega) + Q(\omega)
\xi (\omega) .
\end{align}

According to similar calculations, we find
\begin{align}\label{11}
\delta a_L^\dagger (\omega) =&\ E^*(-\omega) a_{L,\mathrm{in}}^\dagger (\omega) + F^*(-\omega) a_{L,\mathrm{in}} (\omega) \nonumber \\
&\ + G^*(-\omega) a_{R,\mathrm{in}}^\dagger (\omega) + H^* (-\omega) a_{R,\mathrm{in}} (\omega)  \nonumber \\
&\ + Q^*(-\omega) \xi (\omega).
\end{align}

Using the Fourier transform, we obtain
\begin{align}\label{12}
\langle  a_{L,\mathrm{in}} (\omega) a_{L,\mathrm{in}}^\dagger (\omega')\rangle=&\ \frac{1}{\sqrt{2\pi}}\int_{-\infty}^{\infty}\langle a_{L,\mathrm{in}} (t)e^{-i\omega t}dt\nonumber\\
&\ \times\frac{1}{\sqrt{2\pi}}\int_{-\infty}^{\infty} a_{L,\mathrm{in}}^\dagger (t') \rangle e^{-i\omega' t'}dt'\nonumber\\
&\ =\delta(\omega+\omega'),
\end{align}
and
\begin{align}\label{13}
\langle  a_{R,\mathrm{in}} (\omega) a_{R,\mathrm{in}}^\dagger
(\omega')\rangle=&\ \delta(\omega+\omega').
\end{align}

\section{Derivation of Optimal Parameters} \label{appendix C}

According to the quantum-trajectory method~\cite{Plenio98The}, the
non-Hermitian Hamiltonian of the system containing the optical
decay and mechanical damping terms given by~\cite{Plenio98The}
\begin{align}\label{eq:Ke1}
H'=&\ \hbar(\Delta_L-i\frac{\kappa_L}{2}) a_L^{\dagger}a_L + \hbar(\Delta_R' -i\frac{\kappa_R}{2}) a_R^{\dagger}a_R \nonumber\\
&\ +\hbar(\omega_m-i\frac{\gamma_m}{2}) b^{\dagger}b + \hbar J(a_L^{\dagger}a_R + a_R^{\dagger}a_L) \nonumber\\
&\ - \hbar\delta (a_R^{\dagger}a_R)^2 +
i\hbar\epsilon_d(a_L^{\dagger} - a_L) \,,
\end{align}
where $\Delta_R' = \Delta_{R}+\Delta_\mathrm{F}$.

Under the weak-driving conditions, we can make the
ansatz~\cite{Bamba11Origin}:
\begin{align}\label{eq:Ke11}
|\varphi\rangle =&\ C_{00} |0,0\rangle + C_{10} |1,0\rangle +
C_{01} |0,1\rangle + C_{20} |2,0\rangle \nonumber\\&\ + C_{11}
|1,1\rangle + C_{02} |0,2\rangle.
\end{align}
Then we substitute Hamiltonian (\ref{eq:Ke1}) and the general
state (\ref{eq:Ke11}) into the Schr\"{o}dinger equation
\begin{align}\label{eq:Sq}
i\hbar\frac{\mathrm{d}|\varphi\rangle}{\mathrm{d}t} =
H'|\varphi\rangle,
\end{align}
then we have:
\begin{align}\label{eq:Ke111}
H'C_{00} |0,0\rangle=&\ i\hbar\epsilon_d C_{00} |1,0\rangle,\nonumber\\
H'C_{10} |1,0\rangle=&\ \hbar\delta_LC_{10} |1,0\rangle+\hbar J C_{10} |0,1\rangle \nonumber\nonumber\\&\ + i\hbar\epsilon_d C_{10} (\sqrt{2}|2,0\rangle-|0,0\rangle),\nonumber\\
H'C_{01} |0,1\rangle=&\ \hbar\delta_RC_{01} |0,1\rangle+\hbar JC_{01} |1,0\rangle \nonumber\nonumber\\&\ + i\hbar\epsilon_d C_{01} |1,1\rangle,\nonumber\\
H'C_{20} |2,0\rangle=&\ 2\hbar\delta_LC_{20} |2,0\rangle+\sqrt{2}\hbar JC_{20} |1,1\rangle \nonumber\\&\ + i\hbar\epsilon_d C_{20} (\sqrt{3}|3,0\rangle-\sqrt{2}|1,0\rangle),\nonumber\\
H'C_{11} |1,1\rangle=&\ \hbar\delta_LC_{11} |1,1\rangle +\hbar\delta_RC_{11} |1,1\rangle \nonumber\\&\ +\sqrt{2}\hbar JC_{11} (|2,0\rangle+|0,2\rangle) \nonumber\\&\ + i\hbar\epsilon_d C_{11} (\sqrt{2}|2,1\rangle-|0,1\rangle),\nonumber\\
H'C_{02} |0,2\rangle=&\ 2\hbar\delta_RC_{02} |0,2\rangle- 2\delta
C_{02} |0,2\rangle\nonumber\\&\ +\sqrt{2}\hbar JC_{02}
(|1,1\rangle + i\hbar\epsilon_d C_{02} |1,2\rangle,
\end{align}
where the auxiliary functions are $\delta_L =\Delta_L-i\kappa_L/2$
and $\delta_R=\Delta_R'-i\kappa_R/2$, and we have ignored the
effects of the mechanical model, because the phonon states are
decoupled from the photon states [see Eq.~\ref{eq:Ke1}]. By
comparing the coefficients, we have
\begin{align}\label{eq:Ke12}
\frac{\partial C_{00}}{\partial t} =&\ \epsilon_dC_{10},\nonumber\\
i\frac{\partial C_{10}}{\partial t} =&\ \delta_L C_{10} + JC_{01} -\sqrt{2} i \epsilon_d  C_{20},\nonumber\\
i\frac{\partial C_{01}}{\partial t} =&\ (\delta_R- \delta) C_{01} + J C_{10} - i \epsilon_d  C_{11},\nonumber\\
i\frac{\partial C_{11}}{\partial t} =&\ \delta_L C_{11} + (\delta_R - \delta) C_{11} \nonumber\\&\ + \sqrt{2} J (C_{02}+C_{20}) + i \epsilon_d  C_{01},\nonumber\\
i\frac{\partial C_{02}}{\partial t} =&\ 2(\delta_R - \delta) C_{02} + \sqrt{2} J C_{11} -2 \delta C_{02},\nonumber\\
i\frac{\partial C_{20}}{\partial t} =&\ 2(\delta_R- \delta) C_{20}
+ \sqrt{2} J C_{11} +\sqrt{2} i \epsilon_d  C_{10}.
\end{align}

{Then the steady-state coefficients of the 
one- and two-particle states are given as
\begin{align}\label{eq:Sc1}
0=&\ \delta_L C_{10} +  J C_{01} + i\epsilon_d C_{00},\nonumber \\
0=&\ \delta_RC_{01} +  JC_{10},
\end{align}
and
\begin{align}\label{eq:Sc21}
0=&\ 2\delta_L C_{20} +  \sqrt{2} J C_{11} + i\sqrt{2} \epsilon_d C_{10},\nonumber \\
0=&\ (\delta_L + \delta_R )C_{11}
+ \sqrt{2}J C_{20}  + \sqrt{2}J C_{02} + i\epsilon_d C_{01},\nonumber \\
0=&\ 2(\delta_R - \delta)C_{02} +  \sqrt{2} J C_{11},
\end{align}
where we have introduced the dissipative terms
(proportional to $\kappa_L$ and $\kappa_R$) and neglected the
higher-order terms, as justified under the weak-driving
conditions.}

{When we consider $\Delta_L = \Delta_R -\delta
=\Delta$, $\delta=g^2/\omega_m$, $\kappa_L = \kappa_R = \kappa$,
and the condition of $C_{20} = 0$, wehave
\begin{align}\label{eq:Scf1}
0=&\ \kappa^2(2\delta-6\Delta-5\Delta_ \mathrm{F}^2) + 4\Delta^2(2\Delta-2\delta-5\delta\Delta_ \mathrm{F}^2) \nonumber\\
&\ + 4\Delta_ \mathrm{F}(4\Delta\Delta_ \mathrm{F}-3\delta\Delta-\delta\Delta_ \mathrm{F}+\Delta_ \mathrm{F}^2)-4J^2\delta ,\nonumber\\
0 =&\ 8\delta \Delta  - 12 \Delta^2 + \kappa^2 + \Delta_
\mathrm{F}(6\delta - 20 \Delta - 8 \Delta_ \mathrm{F}).
\end{align} }
{By eliminating $\delta$, we obtain}
\begin{align}\label{C13}
a_4\Delta^4 + a_3\Delta^3  +a_2\Delta^2 +a_1\Delta + a_0 = 0,
\end{align}
where
\begin{align}\label{C13-1}
a_0 =&\ \kappa(4J^2-10\Delta_ \mathrm{F}^2)(\kappa^2-8\Delta_ \mathrm{F}^2)-2\kappa(\kappa^4-44\Delta_ \mathrm{F}^4)  ,\nonumber\\
a_1 =& -8\Delta_ \mathrm{F}(6\Delta_ \mathrm{F}^2 \kappa + 10J^2\kappa + 3),\nonumber\\
a_2 =& -8\kappa(2\kappa^2+ 6J^2 +13\Delta_ \mathrm{F}^2)  ,\nonumber\\
a_3 =& -96 \Delta_ \mathrm{F} \kappa ,\nonumber\\
a_4 =& -32 \kappa,
\end{align}
then we find the optimal conditions
\begin{align}\label{C14}
\Delta_\mathrm{opt} \approx&\ \frac{-a_3+\mathrm{sgn}(E) \sqrt{\lambda_1} - \sqrt{\lambda_2}}{4a_4},\nonumber\\
g_\mathrm{opt} =&\ \sqrt{-\frac{\omega_m\left[\Delta_\mathrm{opt}
(4\Delta_\mathrm{opt}^2 + 5\kappa^2 ) + \Delta_ \mathrm{F}
\lambda_3\right]}{2(2J^2-\kappa^2) + 2\Delta_
\mathrm{F}\lambda_{4}}},
\end{align}
where
\begin{align}\label{C14-1}
\lambda_1 =&\ \frac{D +\sqrt[3]{z_1} + \sqrt[3]{z_2}}{3},\nonumber\\ \lambda_2 =&\ \frac{2D - \sqrt[3]{z_1} - \sqrt[3]{z_2} + \sqrt[3] {z_3}}{3},\nonumber\\
\lambda_3 =&\ 20 \Delta_\mathrm{opt}^2 - 8 \Delta_\mathrm{opt}\Delta_ \mathrm{F}-4\Delta_ \mathrm{F}^2+5\kappa^2,\nonumber\\
\lambda_4 =&\ 10\Delta_\mathrm{opt}^2 + 3\Delta_\mathrm{opt} +
2\Delta_\mathrm{F},
\end{align}
and
\begin{align}\label{C14-2}
\mathrm{sgn}(E)=&
\begin{cases}
1 & (E>0) ,\nonumber\\
-1 & (E<0) ,
\end{cases} \\
z_{1,2} =&\ AD + 3 \frac{-B\pm \sqrt{B^2-4 AC}}{2} \,,\nonumber\\
z_3 =&\ D^2- D(\sqrt[3]{z_1} + \sqrt[3]{z_2}) + (\sqrt[3]{z_1} + \sqrt[3]{z_2})^2 - 3A \,,\nonumber\\
A =&\ D^2 - 3F ,\nonumber\\
B =&\ DF - 9E^2 ,\nonumber\\
C =&\ F^2 - 3D E^2,\nonumber\\
D =&\ 3 a_3^2 - 8 a_4 a_2 ,\nonumber\\
E =&\ -a_3^3 + 4 a_4 a_3 a_2 - 8 a_4^2 a_1 ,\nonumber\\
F =&\ 3 a_3^4 + 16 a_4^2 a_2^2 - 16 a_4 a_3^2 a_2 + 16a_4^2a_3a_1 \nonumber\\
&\ - 64a_4^3a_0.
\end{align}

{\bf Funding.} NSF of China under Grants No. 11474087 and No.
11774086, and the HuNU Program for Talented Youth.

\end{document}